\documentclass[manuscript]{emulateapj}
\usepackage{hyperref,amsmath,amsfonts}
\slugcomment{}

\begin{document}

\title{Fermi bubble $\gamma$-rays as a result of diffusive injection of Galactic cosmic rays}
\author{Satyendra Thoudam}
\affil{Department of Astrophysics, IMAPP, Radboud University Nijmegen\\P.O. Box 9010, 6500 GL Nijmegen, The Netherlands}
\date{\today}
\email{s.thoudam@astro.ru.nl}

\begin{abstract}
Recently, the {\it{Fermi}} space telescope has discovered two large $\gamma$-ray emission regions, the so-called ``Fermi bubbles", that extend up to $\sim 50^\circ$ above and below the Galactic center. The $\gamma$-ray emission from the bubbles are found to follow a hard spectrum with no significant spatial variation in intensity and spectral shape. The origin of the emission is still not clearly understood. Suggested explanations include injection of cosmic-ray nuclei from the Galactic center by high-speed Galactic winds, electron acceleration by multiple shocks and stochastic electron acceleration inside the bubbles. In this letter, it is proposed that the $\gamma$-rays can be the result of diffusive injection of Galactic cosmic-ray protons during their propagation through the Galaxy. Considering that the bubbles are slowly expanding, and cosmic rays undergo much slower diffusion inside the bubbles than in the averaged Galaxy and at the same time suffer losses due to adiabatic expansion and inelastic collisions with the bubble plasma, this model can explain the observed intensity profile, the emission spectrum and the measured luminosity without invoking any additional particle production processes unlike other existing models.
\end{abstract}

                  
\section{Introduction}
Recent detailed analysis of the {\it{Fermi}}-LAT data has discovered two giant $\gamma$-ray emission regions extending up to $\sim 50^\circ$ ($\sim 10$ kpc) in Galactic latitude above and below the Galactic center (GC) with a width of $\sim 40^\circ$ in longitude \citep{Su2010}. The $\gamma$-ray emission regions, now popularly known as the ``Fermi bubbles" (FBs), coincide well with the WMAP haze at low latitudes \citep{Fink2004, Dobler2008}, and share their edges with the ROSAT X-ray map \citep{Snowden1997}. More recently, the {\it{Planck}} satellite experiment have shown that the morphology of the FBs is highly consistent with that of the microwave haze \citep{Planck2013}, and also radio measurements have found that the bubbles are coincident with two giant radio lobes that appear to originate from the GC \citep{Carretti2013}. This correlation between the multi-wavelength observations seem to suggest that the FB $\gamma$-rays (measured in the range of $\sim 1-100$ GeV) are produced by high-energy electrons via inverse Compton scattering process, as the same electrons can also simultaneously produce synchrotron radiations in the presence of magnetic fields \citep{Dobler2010}. Moreover, the fact that the FBs are symmetric across the Galactic plane and also centered on the GC encourages to assume the bubbles, and the associated $\gamma$-ray emissions, to have their origin at the GC. 

However, because of severe radiative losses, it would be extremely difficult to transport high-energy electrons from the GC to the far edges of the bubbles. For electrons relevant for producing the FB $\gamma$-rays, convection by Galactic wind would require a wind speed as high as $\sim 10^4$ km s$^{-1}$ which is more than an order of magnitude larger than the typical Galactic wind speed of $\sim 200-300$ km s$^{-1}$, and diffusive transport would need a diffusion coefficient which is $\sim 2-3$ orders of magnitude larger than the standard Galactic value. This problem can be overcome using models based on jet activity of the central active galactic nucleus where jet speeds of over $10^4$ km s$^{-1}$ is readily achievable \citep{Guo2012, Yang2012}. An alternative solution is to consider the production of electrons inside the bubbles itself, for instance, by multiple shock waves generated by periodic star capture by the central supermassive black hole \citep{Cheng2011} or by second-order Fermi acceleration by the plasma wave turbulence present inside the bubbles \citep{Mertsch2011}.
 
On the other hand, hadronic models suffer less constraint at least from the particle injection point of view. \cite{Crocker2011} showed that the FB $\gamma$-rays can be explained if cosmic-ray (CR) nuclei from the GC are injected into the bubbles by fast winds, and if the particles remain trapped inside the bubbles for over $\sim 10^{10}$ yr. In  almost all the models proposed so far (see \citealp{Su2010} for more scenarios), an additional process of high-energy particle production, either inside or outside the bubbles, has been considered. In this letter, we present a simple model which does not invoke any additional sources or particle production processes other than those responsible for the production of bulk of the Galactic CRs. 

In our model, it is assumed that CRs (mainly protons), after leaving their sources, undergo diffusive propagation through the Galaxy. If the FBs are absent of sources, the diffusive streaming of CRs in the direction of density gradient can result into a net flux of CRs injected into the bubbles. In addition, if the bubbles are expanding, there can also be an additional CR injection that scales linearly with the expansion velocity of the bubbles. For uniform injection throughout the bubble surface, the total CR power injected into each bubble (assuming spherical shape) is given by $L=4\pi R^2 u\varepsilon$, where $R$ represents the bubble radius, $u$ the streaming or injection velocity of CRs, and $\varepsilon$ the CR energy density. For CRs streaming at Alfv\'{e}n speed, typically $\sim 100$ km s$^{-1}$, we obtain $L\sim 4\times 10^{40}$ ergs s$^{-1}$ for $R=4.5$ kpc and $\varepsilon=1$ eV cm$^{-3}$, the locally measured energy density. This amount of injected power is $\sim 2$ orders of magnitude larger than the power required in CR protons to produce the measured $\gamma$-ray luminosity of $\sim 2\times 10^{37}$ erg s$^{-1}$ from each bubble. Even if the CR density in the halo is lower by one order of magnitude with respect to the local value, this rough estimate shows that the injection of some fraction of Galactic CRs can easily account for the measured $\gamma$-ray luminosity.

For CRs with an equilibrium number density $N_\mathrm{g}$ in the Galaxy and propagating with diffusion coefficient $D_\mathrm{g}$, their diffusive injection flux into the bubbles is given by $F_{\mathrm{dif}}= D_\mathrm{g}\nabla N_\mathrm{g}\propto D_\mathrm{g} N_\mathrm{g}$, calculated at the edges of the bubbles. Since $N_g$ is related to the CR source spectrum $Q$ as $N_\mathrm{g}\propto Q/D_\mathrm{g}$, we get $F_\mathrm{dif}\propto Q$. Thus, CRs streaming diffusively into the bubbles will follow the same spectral shape as the source CR spectrum in the Galaxy.    

Once injected, CRs undergo diffusive propagation inside the bubbles, while at the same time also convected radially outward by the expanding plasma. If the plasma inside the bubbles is extremely turbulent as suggested by X-ray observations of the Galactic bulge \citep{Yao2007} or if the magnetic field lines are highly tangled \citep{Mcquinn2011}, then the diffusion is expected to be much slower, and the CR transport will be dominated by the convection process. This will result into a CR distribution that peaks towards the bubble edges, in agreement with the  observations \citep{Su2010}. Moreover, for an energy-independent diffusion (which will be justified later), the CR spectrum inside the bubbles will follow the injection spectrum. This is important because for the hadronic origin, the $\gamma$-ray spectrum should mimic that of the parent CRs. The observed $\gamma$-ray spectral index of $\sim 2.0-2.1$ above $\sim 1$ GeV is quite close to the CR source index of $\Gamma\sim 2.1-2.3$ required to produce the locally observed CR spectrum.

Inside the bubbles, secondary electrons (and positrons) can be continuously produced from inelastic collisions of CR protons with the bubble plasma. These electrons can produce synchrotron radiations, and can account for the observed radio and microwave emissions. The production spectrum of such electrons follows the proton spectrum, and has the same index as the source CR spectrum in the Galaxy. If the electrons suffer continuous synchrotron losses, their spectrum at age $t_\mathrm{age}$ of the bubbles will follow $E^{-\Gamma}$ for $E<E_\mathrm{c}$, and $E^{-\Gamma-1}$ for $E>E_\mathrm{c}$, where $E$ denotes the electron energy, and $E_c\propto 1/t_\mathrm{age}$ is the energy at which the energy-loss time equals $t_\mathrm{age}$. Detailed analysis of the combined {\it{Planck}}-WMAP data in the frequency range of $23-61$ GHz have inferred that the electron spectral index in the region is $\sim 2.1$ \citep{Planck2013}. In the present model, this hard spectrum can be explained if the electrons responsible for the microwave emissions have energies less than $E_\mathrm{c}$. 

\section{Model calculations}
\subsection{Cosmic-ray spectrum}
Considering that the bubbles expand outward with constant velocity, the CR proton distribution inside a bubble can be described by a one-dimensional diffusion-loss equation written in the comoving frame of expansion:
\begin{equation}
\frac{\partial}{\partial x}\left(D_\mathrm{b}\frac{\partial N_\mathrm{b}}{\partial x}\right)-\frac{N_\mathrm{b}}{\tau}=\frac{\partial N_\mathrm{b}}{\partial t}
\end{equation}
where $N_\mathrm{b} (x,E,t)$ represents the differential number density of particles with kinetic energy $E$ at time $t$ and position $x$ measured from the bubble boundary with $x>0$ $(<0)$ representing the region inside (outside) the bubble, $D_\mathrm{b}$ represents the diffusion coefficient inside the bubbles, and $\tau=1/(n_\mathrm{b}v\sigma)$ represents the CR inelastic collision time with the bubble plasma of density $n_\mathrm{b}$, $v$ the CR velocity, and $\sigma$ the collision cross-section. Energetic outflows such as unstable large-scale Galactic winds in the inner region of the Galaxy (e.g., \citealp{Everett2008}) can generate a turbulence wave spectrum inside the bubbles that follows $k^{-2}$ in the short wavelength regime $k\gg1/L$, where $k$ denotes the wave number and $L$ is the characteristic length of turbulence injection (e.g., \citealp{Bykov1987}). Then, for turbulence presumably injected at scales of several parsecs which is much larger than the gyro-radii of CRs relevant for the present study, the CR diffusion coefficient is expected to be energy independent. We take $D_\mathrm{b}=K\times 10^{28}$ cm$^2$ s$^{-1}$, where $K$ is a constant. Moreover, since we are mainly interested on CRs with energies above $\sim 1$ GeV, we neglect the very weak energy dependence of $\sigma$ \citep{Kelner2006}, and take a constant value of $\sigma=32$ mb.

For $Q(E)$ spectrum of particles injected at $x=0$ at time $t^\prime$, the solution of Eq. (1) is obtained as,
\begin{align}
N_\mathrm{b}(x,E,t)=&\frac{Q(E)}{\sqrt{\pi D_\mathrm{b}(t-t^\prime)}}\exp\left[\frac{-x^2}{4D_\mathrm{b}(t-t^\prime)}-\frac{(t-t^\prime)}{\tau}\right]
\end{align}
 
The above solution does not include the adiabatic energy loss resulting from the spherical expansion of the bubble. This is taken into account as follows. The adiabatic energy loss rate is given by, 
\begin{equation}
\frac{dE}{dt}=-\frac{2}{3}\frac{E}{t}
\end{equation}
Therefore, particles with energy $E$ at the present time $t$ had energy $E^\prime=E (t/t^\prime)^{2/3}$ at an earlier time $t^\prime$, and distributed within an energy interval $dE^\prime=(t/t^\prime)^{2/3}dE$. Then, the particle distribution in the presence of adiabatic losses can be obtained by replacing $Q(E)\rightarrow Q(E^\prime)(t/t^\prime)^{2/3}$ in Eq. (2). And for continuous injection of particles with flux $F(E)$, the distribution at time $t$ is obtained by writing $Q(E^\prime)\rightarrow F(E^\prime)dt^\prime$, and integrating Eq. (2) over $t^\prime$ as,
\begin{align}
N_\mathrm{b}(x,E,t)=&\frac{1}{\sqrt{\pi D_\mathrm{b}}}\int_0^t dt^\prime\;\frac{F(E^\prime)}{\sqrt{(t-t^\prime)}}\left(\frac{t}{t^\prime}\right)^{2/3}\nonumber\\
&\times\exp\left[\frac{-x^2}{4D_\mathrm{b}(t-t^\prime)}-\frac{(t-t^\prime)}{\tau}\right]
\end{align} 
Note that in the present model, the injection flux  
\begin{equation}
F=\left[D_\mathrm{g}\frac{dN_\mathrm{g}}{dx}+ UN_\mathrm{g}\right]_{x=0}
\end{equation}
where the first term represents the flux of Galactic CRs injected due to their diffusive motion in the Galaxy $F_\mathrm{dif}$, and the second term represents the injection flux due to the expansion of the bubble in the interstellar medium $F_\mathrm{exp}$. Finally, the solution in terms of the radial coordinate $r$ measured from the bubble center can be found by simply replacing $x$ with $R-r$ in Eq. (4).

To estimate the injection fluxes, we write the Galactic CR density in the pure diffusion model as function of perpendicular distance $z$ to the Galactic plane as \citep{Thoudam2008},
\begin{equation}
N_\mathrm{g}(z,E)\propto\frac{Q(E)}{D_\mathrm{g}(E)} f(z)
\end{equation}
where $Q(E)$ represents the source spectrum and $D_\mathrm{g}$ is the diffusion coefficient in the Galaxy. The function
\begin{equation}
f(z)=\int^\infty_0\frac{\sinh[K(H-z)]\times  \mathrm{J_1}(K\Re)dK}{\sinh(KH)\left[K\coth(KH)+\eta v\sigma/(2D_\mathrm{g})\right]}
\end{equation}
has a very weak energy dependence, where $\mathrm{J_1}$ is the Bessel function of order 1, $H$ represents the halo boundary taken to be large enough to contain the FBs, $\Re=20$ kpc the radial size of the source distribution, and $\eta$ the averaged surface density of interstellar gas in the Galactic plane. For $H=10$ kpc, we obtain $D_\mathrm{g}=D_0 (E/3\mathrm{GeV})^{0.6}$ based on the measured boron-to-carbon ratio, where $D_0=6\times 10^{28}$ cm$^2$ s$^{-1}$. 

Then, the CR density gradient along $z$ follows,
\begin{equation}
\frac{dN_\mathrm{g}}{dz}\propto\frac{Q(E)}{D_\mathrm{g}}\frac{df}{dz}
\end{equation}
From Eq. (8), it can be noticed that
\begin{equation}
D_\mathrm{g}\frac{dN_\mathrm{g}}{dz}\propto Q(E)
\end{equation}
which shows that the diffusive injection flux $F_\mathrm{dif}$ follows the source CR spectrum as mentioned before. On the other hand, the injection flux due to the expansion $F_\mathrm{exp}$ follows the ambient CR spectrum, and hence is steeper than $F_\mathrm{dif}$.

To calculated $F_\mathrm{exp}$, we estimate the expansion velocity of the bubble based on the combined {\it{Planck}}-WMAP measurements. Assuming that the electrons radiate at critical frequency, the highest measured frequency of $61$ GHz corresponds to an electron energy of $65.6$ GeV for an estimated magnetic field strength of $B=1.3$ $\mu \mathrm{G}$ inside the bubbles. The latter is calculated at the position of the bubble center (taken at $z=5$ kpc above the GC) using the relation $B(z)=7e^{-z/3\mathrm{kpc}}$ $\mu \mathrm{G}$ used in the {\footnotesize{GALPROP}} CR propagation code \citep{Strong2010}. The fact that {\it{Planck}}-WMAP data has not found any break or steepening in the spectrum up to $61$ GHz implies that the age of the bubbles must be less than the synchrotron loss time of $65.6$ GeV electrons. This gives an upper limit of the bubble age at $t_\mathrm{age}=1.1\times 10^8$ yr. For the present radius of $R=4.5$ kpc for the bubbles, the corresponding lower limit of the expansion velocity is obtained as $U=39.6$ km s$^{-1}$. Choosing a different scale height, say $5$ kpc, will give a slightly larger magnetic field of $\sim 2.5$ $\mu\mathrm{G}$ inside the bubbles, lowering the upper limit of $t_\mathrm{age}$ and increasing the lower limit of $U$ by factor of around 2. These values of magnetic field are less than the estimated equipartition values of $\sim 6$ $\mu\mathrm{G}$ inside the bubbles \citep{Carretti2013}, and $\sim 15$ $\mu\mathrm{G}$ of the bubble walls \citep{Jones2012}.

The CR injection flux might vary for different positions, and also for different directions in the Galaxy, But, for the present study, we neglect such variations and assume uniform injection over the entire bubble surface. We assume the injection flux to correspond to that at $z=5$ kpc as $C\left[F_\mathrm{dif}+F_\mathrm{exp}\right]_{z=5\mathrm{kpc}}$, where we have introduced a constant $C$, hereafter referred to as the injection fraction, in order to take care of the unknown actual fraction of CRs injected. Its value will be determined based on the measured $\gamma$-ray data. If the diffusion coefficient in the Galaxy has a spatial dependence that scales inversely with the magnetic field, we can write $D_0(z)=6\times 10^{28} e^{z/3\mathrm{kpc}}$ cm$^2$ s$^{-1}$ which gives a value of $D_0=31.7\times 10^{28}$ cm$^2$ s$^{-1}$ at $z=5$ kpc. For this value of $D_0$ and $U=39.6$ km s$^{-1}$ obtained above, the injection flux is dominated by $F_\mathrm{dif}$ for CR energies above $1$ GeV. For larger values of $U$, $F_\mathrm{exp}$ will become significant especially at lower energies. This effect will be discussed later in section 3.    

The CR source spectrum is chosen to be a broken power-law with $\Gamma=2.2$ and $2.09$ for energies below and above $300$ GeV. For the assumed form of $D_\mathrm{g}$, this choice of spectral index reproduces well the proton spectrum recently measured by the ATIC \citep{Panov2007}, CREAM \citep{Yoon2011}, and PAMELA \citep{Adriani2011} experiments which exhibit a slight hardening above $\sim 300$ GeV.

\subsection{Gamma-ray emission}
For the proton distribution $N_\mathrm{b}(r,E,t)$, the $\gamma$-ray emissivity $q_\gamma(r,E_\gamma,t)$ of energy $E_\gamma$ is calculated using the inelastic interaction cross-section given by \cite{Kelner2006}. Then, the $\gamma$-ray intensity in a given direction characterized by the Galactic longitude $l$ and the latitude $b$ is calculated as
\begin{equation}
I_\gamma(l,b,E_\gamma,t)=\frac{1}{4\pi}\int^{y_2}_{y_1} q_\gamma(y,E_\gamma,t)dy
\end{equation}
where the integration is performed along the line of sight distance $y$, and the integration limits are determined from the points of intersection of the line of sight with the bubble surface. In Eq. (10), the emissivity previously written as function of $r$ has been carefully expressed in terms of $y$.

\section{Results and discussions}
At any given time, the CR distribution inside the bubbles is governed mainly by the competition between diffusion and convection. Diffusion tends to uniform the distribution while convection (and also, the other subdominant effects: inelastic collision and adiabatic losses) does the opposite. Thus, faster diffusion will produce more uniform distribution due to the increasing CR diffusion distance in the given time.
\begin{figure}
\centering
\includegraphics*[width=0.33\textwidth,angle=-90,clip]{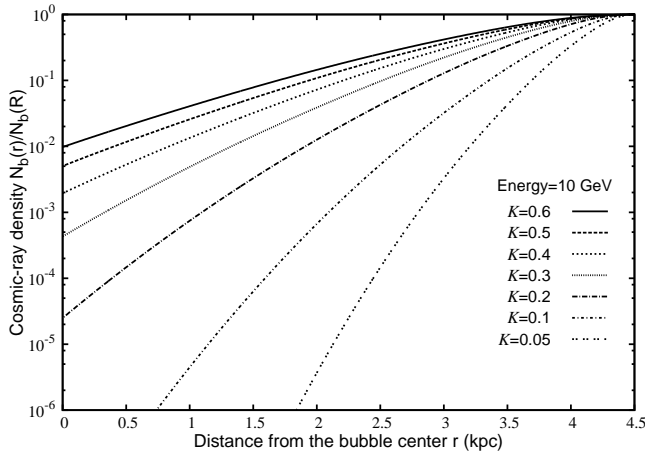}
\caption{\label {fig1} Normalized distribution of $10$ GeV CRs inside a Fermi bubble for different values of $D_\mathrm{b}$ with $K=0.05-0.6$.}
\end{figure}
This is shown in Figure 1 where we have plotted the normalized distributions of $10$ GeV CRs inside a bubble for different values of $D_\mathrm{b}$ taking $K=(0.05-0.6)$. For the calculation (and also, in the following), we take $t=t_\mathrm{age}=1.1\times 10^8$ yr, $U=39.6$ km s$^{-1}$, and $n_\mathrm{b}=3\times 10^{-3}$ cm$^{-3}$. The latter is the averaged value in the region taken from \cite{Everett2008} that explains the diffuse soft X-ray emission. It can be seen that the distribution becomes flatter towards the bubble center with increasing $K$. The distribution is expected to be similar at all energies because of the energy-independent nature of $D_\mathrm{b}$.

\begin{figure}
\centering
\includegraphics*[width=0.33\textwidth,angle=-90,clip]{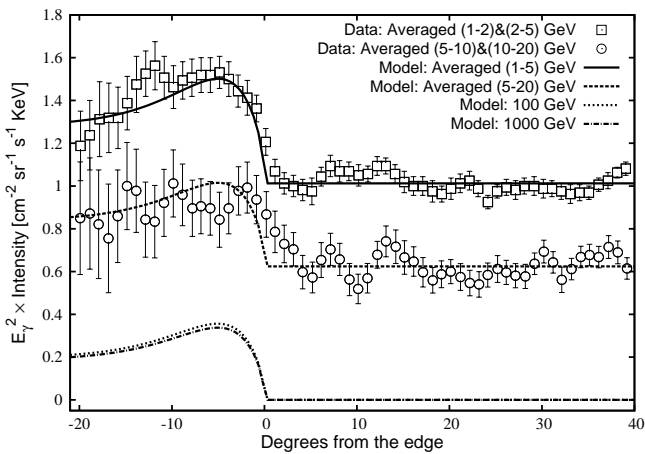}
\caption{\label {fig1} Projected $\gamma$-ray intensity profile of the Southern bubble for the averaged $1-5$ GeV (solid line) and $5-20$ GeV (dashed line). Also shown are the predictions for $100$ GeV (dotted line) and $1000$ GeV (dot-dashed line) energies. Data: \cite{Su2010}.}
\end{figure}

Once the values of $t$, $U$ and $n_\mathrm{b}$ are fixed, the CR distribution inside the bubbles is determined by the choice of $D_\mathrm{b}$. Here, its value is chosen such that the resulting projected $\gamma$-ray intensity distribution matches the measured  profile. It is found that choosing $K=0.26$ produces a good fit to the measured data as shown in Figure 2, where the data corresponds to the Southern bubble. The model predictions for both the averaged $1-5$ GeV (solid line) and $5-20$ GeV (dashed line) are added with backgrounds obtained by fitting horizontal lines to the respective data between $5^\circ$ and $40^\circ$. It can be mentioned that neither the hadronic model presented in \cite{Crocker2011} nor the leptonic model based on diffusive shock acceleration \citep{Cheng2011} can satisfactorily explain the measured sharp edges shown in Figure 2. Both these models predicted a constant volume emissivity throughout the bubbles which will produce softer edges on the projected profile. Also shown in Figure 2 are the predictions for $100$ GeV (dotted line) and $1000$ GeV (dot-dashed line) $\gamma$-rays which can be tested in future. As expected, their profiles look very similar to those at low energies. Our result at high energies, say at $1000$ GeV, is clearly different from that expected from the leptonic stochastic acceleration model presented in \cite{Mertsch2011} which predicted a significant edge brightening at high energies.  

\begin{figure}
\centering
\includegraphics*[width=0.33\textwidth,angle=-90,clip]{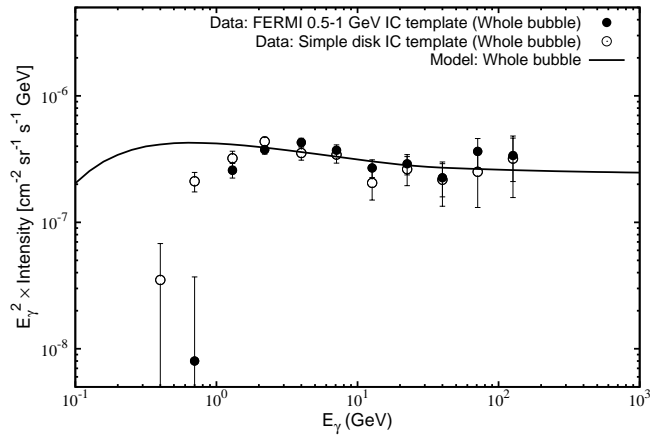}\\
\includegraphics*[width=0.33\textwidth,angle=-90,clip]{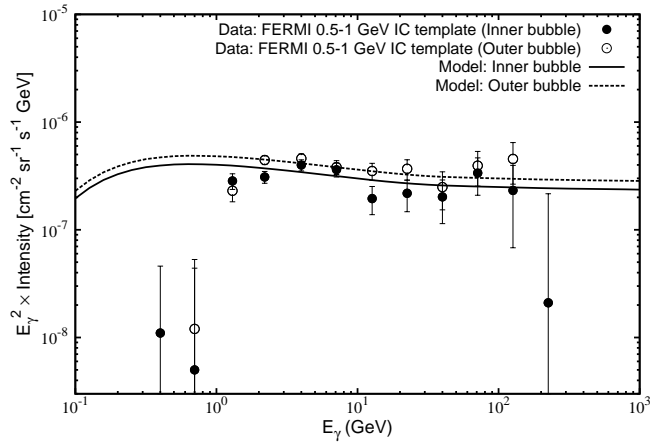}
\caption{\label {fig1} Gamma-ray spectra for a whole bubble (top), and for the bubble inner and outer regions (bottom). Data: \cite{Su2010}.}
\end{figure}

The $\gamma$-ray spectrum averaged over a whole bubble is shown in Figure 3 (top). The calculation assumes an injection fraction of $C=0.8$ which is also the same injection value used in Figure 2. The model prediction is found to be in good agreement with the data in the energy range of $\sim 1-100$ GeV where the measurement uncertainties are small. It is interesting to see that the same source index required to explain the measured CR spectrum in the pure diffusion model also reproduces the FB $\gamma$-ray spectrum. This would be difficult if one uses the re-acceleration model of CR propagation which requires a steeper source index of $\sim 2.4$. 

Spectra for the inner and outer regions of a bubble separated as in \cite{Su2010} are shown in Figure 3 (bottom). The outer region is taken as a shell with thickness $1$ kpc. Our model predictions are in good agreement with the measured data between $\sim 1-100$ GeV, and reproduce the measured spectral uniformity in the two regions. Even the slightly higher intensity measured in the outer region seems to be explained.

\begin{figure}
\centering
\includegraphics*[width=0.33\textwidth,angle=-90,clip]{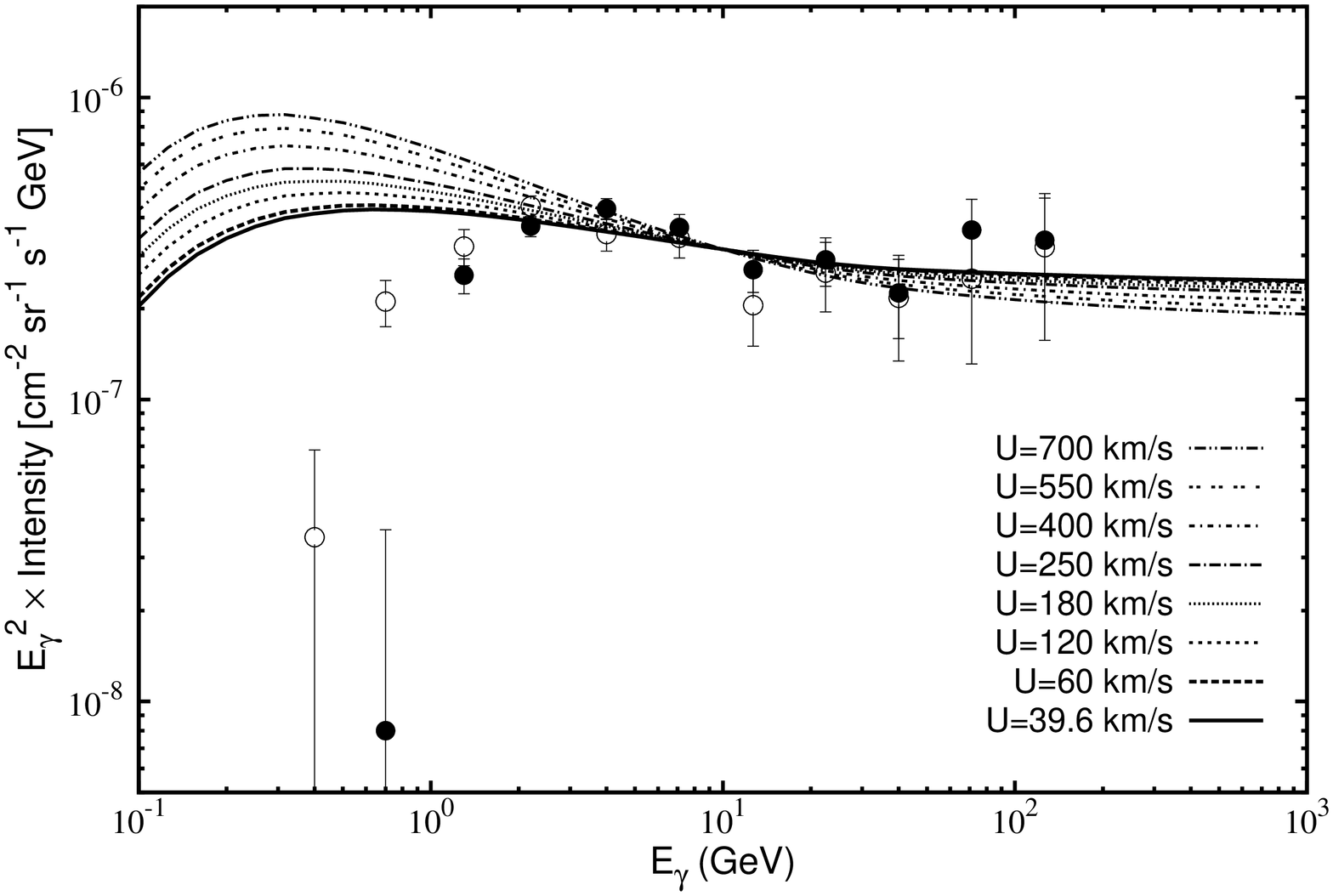}\\
\includegraphics*[width=0.33\textwidth,angle=-90,clip]{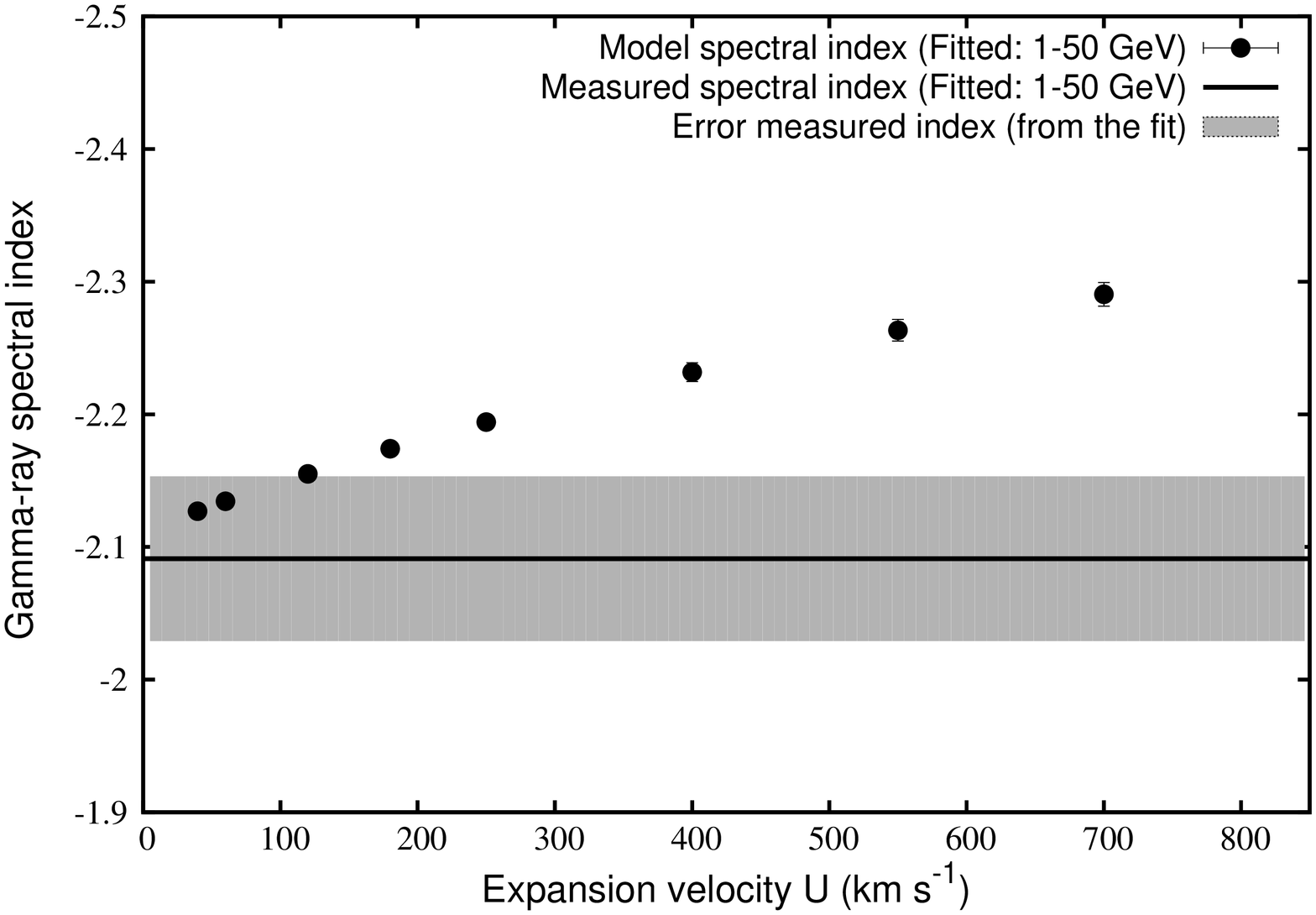}
\caption{\label {fig1} Top: Gamma-ray spectra for a whole bubble for different values of expansion velocity $U=39.6-700$ km s$^{-1}$. Bottom: Fitted spectral indices for the spectra shown in the upper panel as function of expansion velocity, compared with the measured index.}
\end{figure}

The results presented above consider the lower limit of expansion velocity inferred from the combined {\it{Planck}}-WMAP data. Choosing higher expansion velocity will increase the contribution of $F_\mathrm{exp}$, thereby making the total injection spectrum steeper. This will result into a steeper $\gamma$-ray spectrum. This is shown in Figure 4 (top) for different velocities in the range of $U=39.6-700$ km s$^{-1}$, where the spectra are normalized to that of  $U=39.6$ km s$^{-1}$ at $10$ GeV. It can be noticed that the model spectrum becomes steeper as $U$ increases. This is shown explicitly in Figure 4 (bottom), where the spectral indices for the different results are plotted as function of $U$. The indices (represented by points) are obtained by fitting the model spectra between $1-50$ GeV. The solid line together with the shaded region represents the measured index of $-2.09\pm 0.06$ obtained by fitting the measured spectrum in the same energy range. It can be noticed that already at $U=180$ km s$^{-1}$, the model prediction becomes inconsistent with the measured index. This sets an upper limit on the expansion velocity at $U<180$ km s$^{-1}$, which translates into a lower limit of the bubble age at $t_{\mathrm{age}}>2.44\times 10^7$ yr. Consequently, this implies a synchrotron break in the electron spectrum at energy below $298$ GeV, and also a corresponding break in the synchrotron emission spectrum at frequency below $1258$ GHz which can be checked by future measurements. 

For typical sound speed of $\sim 100$ km s$^{-1}$ (for gas temperature of $\sim 10^6$ K) in the Galactic halo, the limit of $U<180$ km s$^{-1}$ implies a Mach number $M<1.8$, even less at $M<1.2$ if the CR pressure is also taken into account. Such weak shocks are known to be rather inefficient for particle acceleration, justifying our neglect of CR acceleration by the expanding bubbles. Moreover, for a Galactic gravitational acceleration value of $\sim 10^{-8}$ cm s$^{-2}$ in the halo \citep{Ferriere1998}, this upper limit of $U$ gives an extend of the plasma flow to $\sim 6-7$ kpc, roughly the size of the FBs. In addition, the present values of $n_\mathrm{b}$ and $B$ for the bubbles give an Alfv\'{e}n speed of $52$ km s$^{-1}$ which corresponds to a plasma beta value of $\beta \sim 4$, indicating a low magnetic pressure  bubbles.

\section{Conclusions}
We have shown that the $\gamma$-rays from the Fermi bubbles can be a result of diffusive injection of Galactic CR protons during their propagation through the Galaxy. Some important observed properties of the bubbles are explained. Unlike other existing models, our proposed model does not consider any additional particle production processes or sources other than those responsible for the production of Galactic CRs.

\acknowledgements{I wish to thank J\"org Rachen for very detailed and insightful discussions on the model. I also thank David Jones for various helpful conversations on the topic.}

\end{document}